\documentclass[12pt]{article}
\usepackage{natbib}
\usepackage{amssymb}
\usepackage{amsmath}
\usepackage{graphicx}

\begin{document}

\noindent AN EMPIRICAL COMPARISON OF GLOBAL AND LOCAL FUNCTIONAL DEPTHS\footnote{This research was partially supported by Spanish Ministry grant ECO2015-66593-P.}

\vskip 5mm
\noindent Carlo Sguera

\noindent UC3M-Banco Santander Institute on Financial Big Data

\noindent Universidad Carlos III de Madrid

\noindent Madrid, Getafe, Spain

\noindent carlo.sguera@uc3m.es

\vskip 5mm
\noindent Rosa E. Lillo

\noindent Department of Statistics; UC3M-Banco Santander Institute on Financial Big Data

\noindent Universidad Carlos III de Madrid

\noindent Madrid, Getafe, Spain

\vskip 3mm

\noindent ABSTRACT

A functional data depth provides a center-outward ordering criterion which allows the definition of measures such as median, trimmed means, central regions or ranks in a functional framework. A functional data depth can be global or local. With global depths, the degree of centrality of a curve $x$ depends equally on the rest of the sample observations, while with local depths, the contribution of each observation in defining the degree of centrality of $x$ decreases as the distance from $x$ increases. We empirically compare the global and the local approaches to the functional depth problem focusing on three global and two local functional depths. First, we consider two real data sets and show that global and local depths may provide different insights. Second, we use simulated data to show when we should expect differences between a global and a local approach to the functional depth problem.

\vskip 4mm

\section{Introduction}
\label{sec:intro}
A functional sample is a collection of $n$ functions or curves $x_{1} = x_{1}(t), \ldots, x_{n} = x_{n}(t)$ defined on a compact interval of the real line. In practice, the elements of any functional sample have an idiosyncratic measurement limitation since they can only be measured on discrete grids, say $t_{1}, \ldots, t_{N}$. However, for insightful analyses, it is often convenient to take into account their underlying infinite-dimensional and functional natures. Indeed, nowadays the theory of statistics for functional data is a well established field with a great amount of applications and ongoing research. See, for instance, \cite{RamSil2005}, \cite{HorKok2012}, \cite{Cue2014} and \cite{kokoszka2017introduction} for overviews of Functional Data Analysis (FDA).\\
\indent In this paper we deal with the notion of data depth in the functional framework. A functional depth provides a center-outward data ordering criterion that, for example, allows the definition of the functional median or ranks. Moreover, the data ordering provided by a functional depth can be exploited in other statistical tasks such as the computation of trimmed means (\citeauthor{FraMun2001}, \citeyear{FraMun2001}), the construction of central regions (e.g., \citeauthor{SunGen2011}, \citeyear{SunGen2011}; \citeauthor{HynSha2010}, \citeyear{HynSha2010}; \citeauthor{narisetty2016extremal}, \citeyear{narisetty2016extremal}), supervised classification (e.g., \citeauthor{cuevas2007robust}, \citeyear{cuevas2007robust}; \citeauthor{SguGalLil2014}, \citeyear{SguGalLil2014}; \citeauthor{MosMoz2015}, \citeyear{MosMoz2015}; \citeauthor{cuesta2017hbox}, \citeyear{cuesta2017hbox}; \citeauthor{hubert2017multivariate}, \citeyear{hubert2017multivariate}), outlier detection (e.g., \citeauthor{FebGalGon2007}, \citeyear{FebGalGon2007}, \citeyear{FebGalGon2008}; \citeauthor{ArrRom2014}, \citeyear{ArrRom2014}; \citeauthor{HubEtAl2015}, \citeyear{HubEtAl2015}; \citeauthor{SguGalLil2015}, \citeyear{SguGalLil2015}; \citeauthor{narisetty2016extremal}, \citeyear{narisetty2016extremal}; \citeauthor{azcorra2018unsupervised}, \citeyear{azcorra2018unsupervised}), principal component analysis (\citeauthor{shang2014survey}, \citeyear{shang2014survey}) or profile monitoring (\citeauthor{GueVar2015}, \citeyear{GueVar2015}), among others.\\
\indent Behind any implementation of the idea of data depth there is an explicit or implicit approach to the depth problem. Particularly, in the multivariate framework, where the notion of depth originated, we find global and local depths. A multivariate global depth provides a data ordering based on the behavior of each observation relative to the complete sample. Several implementations of this notion have been proposed in the literature, e.g., \cite{Tuk1975} proposed the halfspace depth, \cite{Liu1990} introduced the simplicial depth, while \cite{Ser2002} defined the spatial depth.\\
\indent On the contrary, a multivariate local depth provides a data ordering based on the behavior of each observation relative to a certain neighborhood. Regarding existing multivariate local depths, \cite{CheDanPenBar2009} proposed a local version of the spatial depth, the kernelized spatial depth, in the context of local outlier detection, that is, the search of observations that are atypical relative to their neighborhood and not only to the whole data set. Using bivariate motivating examples, \cite{CheDanPenBar2009} show that the kernelized spatial depth captures well the structure of data clouds when distributions deviate from being spherical and symmetric, while its global counterpart does not. Moreover, \cite{AgoRom2011} introduced local versions of the halfspace and simplicial depths to provide tools for recording the local space geometry near any observation and to satisfactorily deal with, for example, multimodal data sets. Finally, \cite{PaiVan2013} defined the $\beta$-local depth to analyze data sets generated from distributions that might be multimodal or have a nonconvex support. Therefore, multivariate local depths focus on data that have some complex or local features, and they usually capture better the underlying structure of data in such nonstandard scenarios.\\
\indent Also in FDA the depth problem can be dealt with a global or a local approach. \cite{nieto2016topologically} and \cite{gijbels2017general} proposed desirable properties of functional data depths. More precisely, \cite{nieto2016topologically} provided a formal definition of functional data depth on the basis of properties related with curve topological features such as continuity, smoothness or contiguity, while \cite{gijbels2017general} showed that the conditions needed for some properties developed by \cite{nieto2016topologically} were extremely demanding and consequently these authors established new properties that are more easily met by common functional depths. These properties are of great theoretical and practical significance, but they do not provide researchers and practitioners with guidance regarding differences between global and local functional depths. Therefore, the main aim of this paper is to point out the structural differences between global and local functional data depths and to illustrate that local functional data depths are better suited for capturing the underlying structure of functional samples with complex features such as the presence of multimodality and the presence of atypical curves. The analysis presented in the paper is expected to help users to decide in which practical situations it is convenient to use functional data depths of one kind or another. This analysis is planned to complement previous main findings in other papers that showed that local functional data depths provide with powerful methods to perform supervised classification (\citeauthor{SguGalLil2014}, \citeyear{SguGalLil2014}) and outlier detection (\citeauthor{CueFebFra2006}, \citeyear{CueFebFra2006}; \citeauthor{SguGalLil2015}, \citeyear{SguGalLil2015}).\\
\indent As global oriented depths, in this paper we consider the Fraiman and Muniz depth (\citeauthor{FraMun2001}, \citeyear{FraMun2001}), which measures how long a curve remains in the middle of a sample of functional observations, the modified band depth (\citeauthor{LopRom2009}, \citeyear{LopRom2009}), which is based on a measure of how much a curve is inside the bands defined by all the possible pairs of curves of a sample, and the functional spatial depth (\citeauthor{ChaCha2014b}, \citeyear{ChaCha2014b}), which is a functional version of the multivariate spatial depth. As local-oriented depths, we consider the h-modal depth (\citeauthor{CueFebFra2006}, \citeyear{CueFebFra2006}), which measures how densely a curve is surrounded by other curves in a sample, and the kernelized functional spatial depth (\citeauthor{SguGalLil2014}, \citeyear{SguGalLil2014}), which represents an explicit local version of the functional spatial depth.\\
\indent We motivate this study by means of two real data examples that involve the presence of some functional local features such as bimodality, presence of isolated observations and potential outliers, or asymmetry. Then, we use a simulation study to analyze the behavior of global and local depths under the presence of complex features, and we observe that global and local functional depths may provide fairly different data insights in specific scenarios.\\
\indent The remainder of the article is organized as follows. In Section \ref{sec:motExa} we recall the definitions of the five functional depths under study and use two real data sets as motivating examples to show that global and local depths may behave differently. In Section \ref{sec:stu} we present the results of a simulation study designed to understand when different behaviors between global and local depths should be expected. In Section \ref{sec:rea} we recover the two real data sets presented in Section \ref{sec:motExa} to corroborate our synthetic results. Finally, in Section \ref{sec:con} we draw some conclusions.

\section{Comparing global and local depths: motivating examples}
\label{sec:motExa}
We open this section presenting the definitions of the functional depths that we compare in this paper, and in particular their empirical versions. Since functional data are in practice observed at a discretized set of points, note that the implementations of the definitions that we present all involve an approximation step\footnote{Implementations of the functional depths considered in this paper are available in the \textit{fda.usc} and \textit{depthTools} R packages.}.

\subsection{Functional depths}
\label{ssec:FD}

\cite{FraMun2001} introduced the first implementation of the notion of depth for functional data, and their idea basically consists in integrating univariate depths. Let $\mathbb{H}$ be an infinite-dimensional Hilbert space. The Fraiman and Muniz depth of $x \in \mathbb{H}$ with respect to the functional sample $Y_{n} = \left\{y_{1}, \ldots, y_{n}\right\}$ is defined as

\begin{equation}
\label{eq:FMD}
FMD(x, Y_{n}) = \int_{I} D_{u}(x(t), Y_{n}(t)) dt,
\end{equation}

\noindent where $I \in \mathbb{R}$ is the interval where $x$ and the elements of $Y_{n}$ are observed, $D_{u}(\cdot, \cdot)$ is any univariate depth (i.e., the univariate version of a multivariate depth), $x(t)$ is the value of $x$ at $t \in I$, and $Y_{n}(t)$ is the vector composed of the $n$ values $y_{1}(t), \ldots, y_{n}(t)$.\\
\indent \cite{LopRom2009} proposed the modified band depth (MBD), which is based on the graphical representation of functional data and on the bands defined by pairs of curves. The MBD of $x$ with respect to $Y_{n}$ is given by

\begin{equation}
\label{eq:MBD}
MBD(x, Y_{n}) = {{n}\choose{2}}^{-1} \sum_{i=1}^{n-1} \sum_{j=i+1}^{n} \lambda_{r}\left(A(x; y_{i}, y_{j})\right),
\end{equation}

\noindent where

\begin{equation}
\label{eq:A}
A(x; y_{i}, y_{j}) = \left\{t \in I: \min\limits_{r = i, j} y_{r}(t) \leq x(t) \leq \max\limits_{r = i, j} y_{r}(t)\right\},\ i, j = 1, \ldots, n.
\end{equation} 

\noindent and $\lambda_{r}\left(A(x; y_{i}, y_{j})\right) = \lambda\left(A(x; y_{i}, y_{j})\right)/\lambda(I)$, while $\lambda$ is the Lebesque measure on $\mathbb{R}$. Note that MBD is known to be closely related to FMD: for more details on this aspect, see for example \cite{nieto2016topologically}.\\
\indent \cite{ChaCha2014b} introduced an extension of the multivariate spatial depth, the functional spatial depth (FSD), with the aim of considering the geometry of the data to assign depth values. The FSD of $x$ with respect to $Y_{n}$ is defined as 

\begin{equation}
\label{eq:fsd}
FSD(x, Y_{n}) = 1 - \frac{1}{n} \left\|\sum_{i=1; y_{i} \neq x}^{n}\frac{x - y_{i}}{\|x - y_{i}\|}\right\|,
\end{equation}

\noindent where $\|\cdot\|$ is the norm induced by the inner product $\langle \cdot, \cdot \rangle$ defined in $\mathbb{H}$.\\
\indent \cite{CueFebFra2006} extended the concept of mode to the functional setup, and as by-product defined the h-modal depth (HMD). The HMD of $x$ with respect to $Y_{n}$ is given by

\begin{equation}
\label{eq:HMD}
HMD(x, Y_{n}) = \frac{1}{n} \sum_{i=1}^{n} \kappa(x, y_{i}),
\end{equation}

\noindent where $\kappa: \mathbb{H} \times \mathbb{H} \rightarrow \mathbb{R}$ is a kernel function depending on a bandwidth $h$.\\
\indent Finally, \cite{SguGalLil2014} introduced the kernelized functional spatial depth modifying FSD in the following way:

\begin{equation}
\label{eq:phi_kfsd}
KFSD(x, Y_{n}) = 1 - \frac{1}{n} \left\|\sum_{i=1}^{n}\frac{\phi(x) - \phi(y_{i})}{\|\phi(x) - \phi(y_{i})\|}\right\|,
\end{equation}

\noindent where $\phi: \mathbb{H} \rightarrow \mathbb{F}$ is an embedding map and $\mathbb{F}$ is a feature space. Since $\phi$ can be defined implicitly by a positive definite and stationary kernel through $\kappa(x, y) = \langle \phi(x), \phi(y)\rangle,\ x, y \in \mathbb{H}$, and after some standard calculations, the kernel-based definition of KFSD is given by

{\footnotesize
\begin{equation}
\label{eq:kfsd}
KFSD(x, Y_{n}) = 1 - \frac{1}{n} \sqrt{\left(\sum_{\substack{i,j=1; \\ y_{i} \neq x; y_{j} \neq x}}^{n}\frac{\kappa(x,x)+\kappa(y_{i},y_{j})-\kappa(x,y_{i})-\kappa(x,y_{j})}{\sqrt{\kappa(x,x)+\kappa(y_{i},y_{i})-2\kappa(x,y_{i})}\sqrt{\kappa(x,x)+\kappa(y_{j},y_{j})-2\kappa(x,y_{j})}}\right)}.
\end{equation}
}

Note that the pair FSD-KFSD represents the unique case where one functional depth (KFSD) is a direct local version of another (FSD), and therefore in what follows we briefly focus on this pair to give more details about the juxtaposition between the global and the local approaches to the depth problem.\\
\indent On the one hand, $FSD(x, Y_{n})$ depends equally on all the possible deviations of $x$ from $y_{i}, \mbox{ for } i = 1, \ldots, n$. Therefore, behind FSD there is an approach based on the following fundamental assumption: each $y_{i}$ should count equally in defining the degree of centrality of $x$. This is the feature that turns FSD into a global-oriented functional depth. A similar approach is behind FMD and MBD.\\
\indent On the other hand, as a modification of FSD, KFSD aims to substitute the equally dependence of the depth value of $x$ on the $y_{i}$'s with a kernel-based dependence producing that a $y_{i}$ closer to $x$ influences more the depth value of $x$ than a $y_{i}$ that is more distant. Therefore, the alternative approach behind KFSD suggests that the contribution of each $y_{i}$ in defining the degree of centrality of $x$ should decrease for $y_{i}$'s distant from $x$. This is the feature that turns KFSD in a local-oriented functional depth. A similar approach is behind HMD.\\
\indent Moreover, the choice of the kernel makes KFSD and HMD extremely flexible tools as it allows the practitioner to implement her/his preferences about the form of the neighborhoods of $x$. Additionally, the kernel bandwidth allows to tune the size of the neighborhood of $x$. In this paper we implement KFSD and HMD using a Gaussian kernel and setting the kernel bandwidth equal to the 25\% percentile of the empirical distribution of {\footnotesize$\left\{\|y_{i}-y_{j}\|, i = 1, \ldots, n; i < j \leq n\right\}$}. Such a bandwidth defines fairly local versions of KFSD and HMD that will be compared to global depths such as FSD, FMD and MBD.\\
\indent The rest of this section is devoted to two motivating examples that consider real functional data sets. First, the phonemes data set, analyzed previously by \cite{HasBujTib1995}, \cite{FerVie2006} or \cite{SguGalLil2014}, among others. Second, the NO$_{x}$ data set, studied also by \cite{FebGalGon2007}, \cite{FebGalGon2008}, \cite{HorKok2012} or \cite{SguGalLil2015}, among others.

\subsection{Phonemes data}
\label{ssec:motExa02}
\begin{figure*}[!htbp]
\centering
\includegraphics[width=0.75\textwidth]{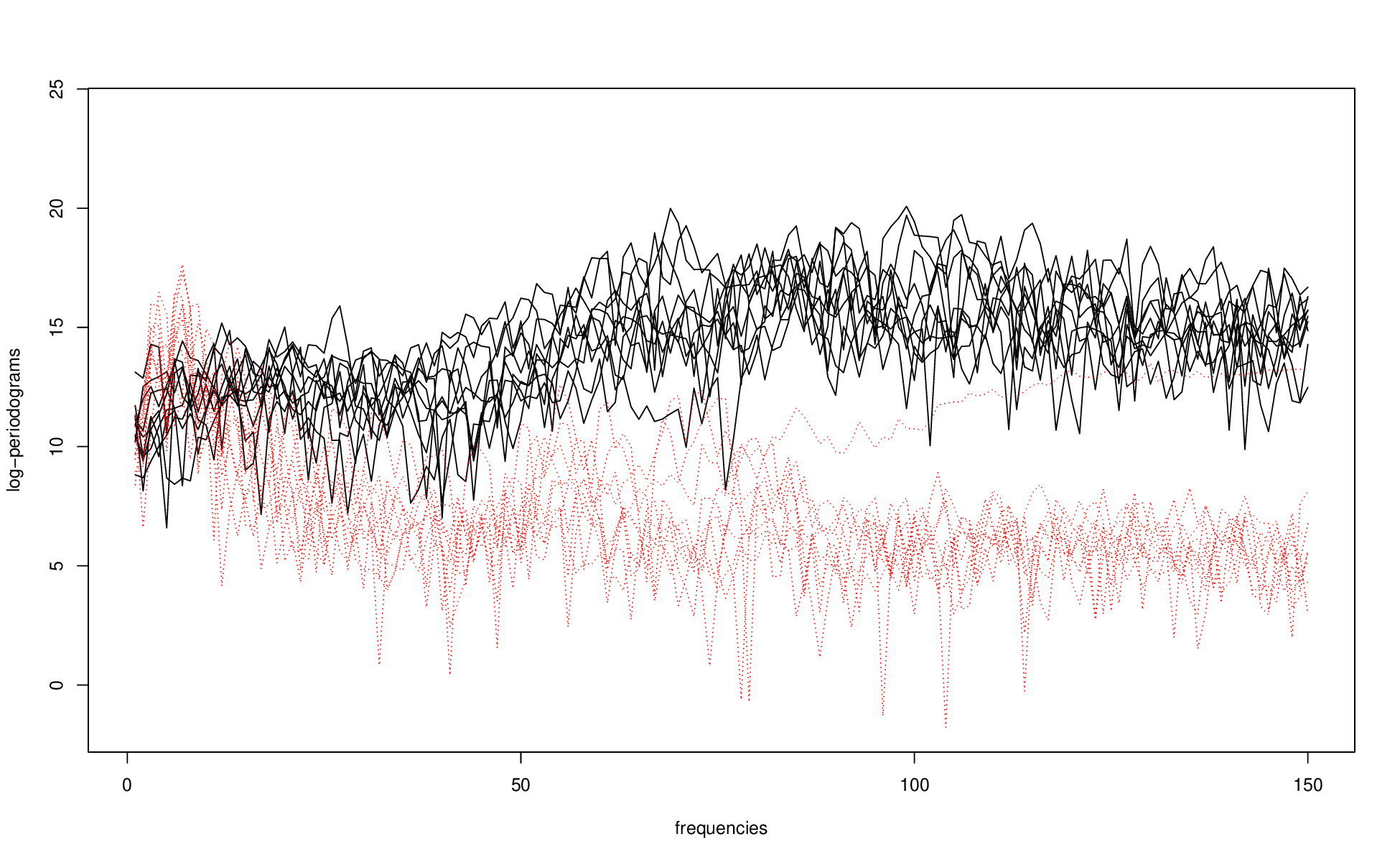}
\caption{Log-periodograms of 10 speakers pronouncing \textit{sh} (solid curves) and of 10 speakers pronouncing \textit{dcl} (dotted curves).}
\label{fig:somePhonemes}
\end{figure*}

The phonemes data set, available in the \textit{fda.usc} R package (\citeauthor{FebOvi2012}, \citeyear{FebOvi2012}), consists in log-periodograms corresponding to recordings of speakers pronouncing specific phonemes. A detailed description of the data set which contains information about five speech frames corresponding to five phonemes can be found in \cite{FerVie2006}. In this subsection we consider 50 observations for the phoneme \textit{sh} as in \underline{sh}e and 50 observations for the phoneme \textit{dcl} as in \underline{d}ark. For illustrative purposes, Figure \ref{fig:somePhonemes} shows 10 randomly chosen log-periodograms for each phoneme. As in \cite{FerVie2006}, we consider the first 150 frequencies from each recording.

\begin{figure*}[!htbp]
\centering
\includegraphics[width=0.75\textwidth]{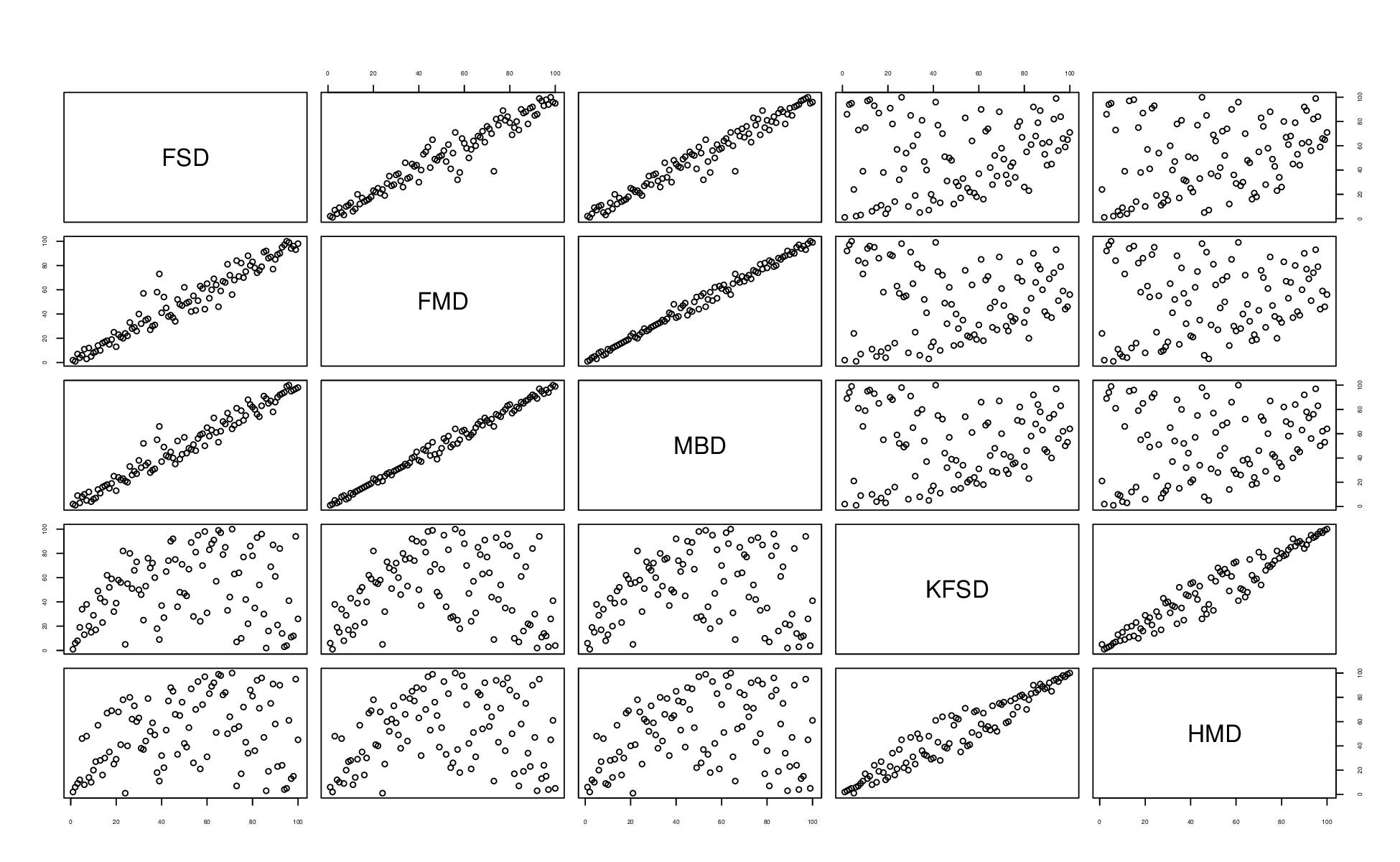}
\caption{Scatter plots of the ten possible pairs of depth-based ranks for the phonemes data set.}
\label{fig:phonemes_Pairs}
\end{figure*}

Treating the above described data set as a unique sample, we obtain data that show bimodality, in particular starting from frequencies around 40, and a central region where fall few isolated observations (we see one of them in Figure \ref{fig:somePhonemes}). Our first goal is to show that global and local depths may behave differently in presence of such complex data features, and since the center-outward ordering of curves is possibly the main by-product of any depth analysis, we evaluate a depth measure considering the ranks associated to its values\footnote{Note that the higher the depth values, the higher the associated ranks.}. We first consider all the possible pairs of depths, and then we focus on the pair FSD-KFSD due to their direct relationship.\\
\indent Figure \ref{fig:phonemes_Pairs} shows the scatter plots of the ten possible pairs of depth-based ranks, and we observe strong relationships between either global or local depths, and relatively weaker relationships between global and local depths. 

In Table \ref{tab:phonemes_Spe} we report the Spearman's rank correlation coefficients corresponding to Figure \ref{fig:phonemes_Pairs}, and they confirm the visual inspection of the figure: the coefficients are never less than 0.96 between either global or local depths, while they are never greater than 0.26 between a global and a local depth.

\begin{table}[!htbp]
\caption{Spearman's rank correlation matrix of FSD, FMD, MBD, KFSD and HMD values for the phonemes data set.}
\centering
\scalebox{1}{
\begin{tabular}{c|ccccc}
\hline
 & FSD & FMD & MBD & KFSD & HMD\\
\hline
FSD & 1.00 & 0.97 & 0.98 & 0.17 & 0.26\\
FMD & 0.97 & 1.00 & 0.99 & 0.02 & 0.13\\
MBD & 0.98 & 0.99 & 1.00 & 0.08 & 0.19\\
KFSD & 0.17 & 0.02 & 0.08 & 1.00 & 0.96\\
HMD & 0.26 & 0.13 & 0.19 & 0.96 & 1.00\\
\hline
\end{tabular}}
\label{tab:phonemes_Spe}
\end{table}

\begin{figure*}[!htbp]
\centering
\includegraphics[width=0.75\textwidth]{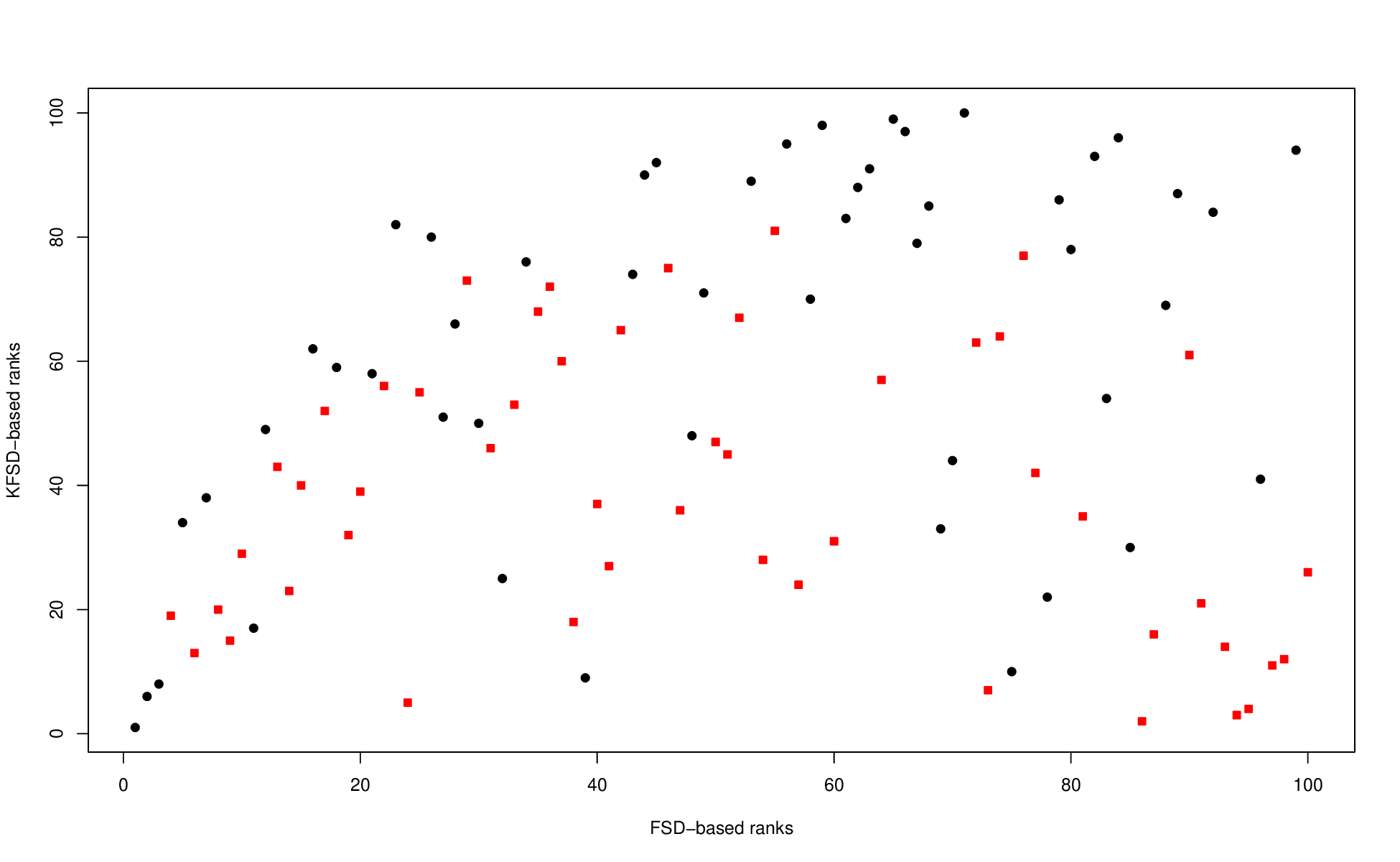}
\caption{Scatter plot of the FSD-based and KFSD-based ranks. The \textit{sh}'s ranks are represented as circles and the \textit{dcl}'s ranks are represented as squares.}
\label{fig:phonemes_RR}
\end{figure*}

In Figure \ref{fig:phonemes_RR} we focus on the scatter plot of the FSD-based and KFSD-based ranks to compare more in detail the behaviors of a global and a local depth: it is clear that there are important differences in terms of ranks, except for low FSD-based ranks (lower than 20). 

Therefore, the bimodal functional phonemes data set represents a clear example of global and local depths showing very different behaviors. After the simulation study of Section \ref{sec:stu}, we recover this data set in Section \ref{sec:rea} for a further analysis.


\subsection{Nitrogen oxides (NO$_{x}$) data}
\label{ssec:motExa01}
\begin{figure*}[!htbp]
\centering
\includegraphics[width=0.75\textwidth]{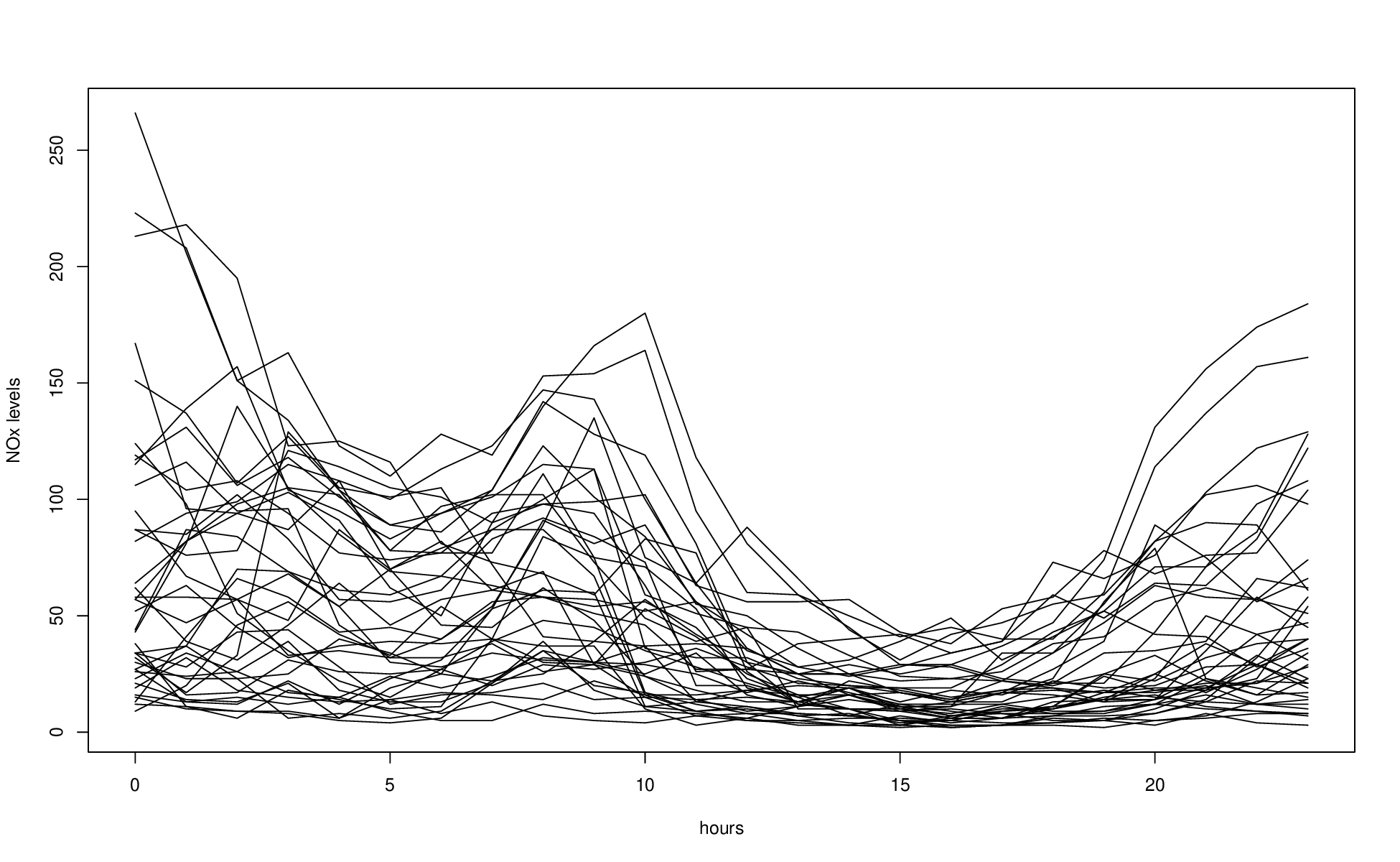}
\caption{NO$_{x}$ levels (nanograms per cubic meter) measured every hour of 39 nonworking days between 23/02/2005 and 26/06/2005 close to an industrial area in Poblenou, Barcelona, Spain.}
\label{fig:noxNW}
\end{figure*}

The nitrogen oxides (NO$_{x}$) data set, also available in the \textit{fda.usc} R package, consists in the nitrogen oxides (NO$_{x}$) emission daily levels measured in a Barcelona area between 2005-02-23 and 2005-06-29. More details about this data set can be found in \cite{FebGalGon2008}, where it is used to implement functional outlier detection techniques after splitting the whole data set in two samples, one containing curves referring to working days, the other to nonworking days. In this subsection we consider the 39 curves of the nonworking days, which are shown in Figure \ref{fig:noxNW}. 

\begin{figure*}[!htbp]
\centering
\includegraphics[width=0.75\textwidth]{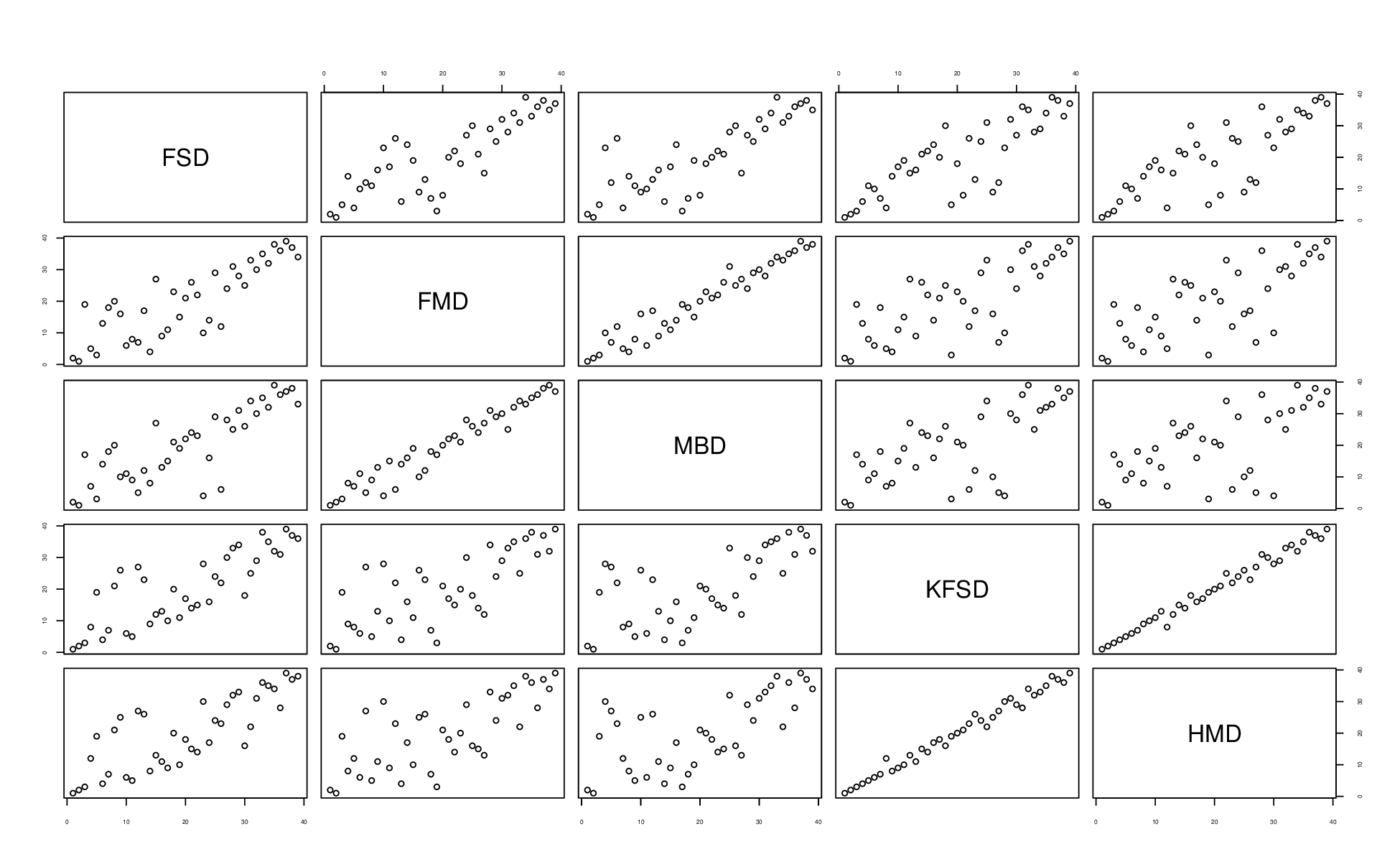}
\caption{Scatter plots of the ten possible pairs of depth-based ranks for the NO$_{x}$ data set.}
\label{fig:nox_Pairs}
\end{figure*}

\begin{figure*}[!htbp]
\centering
\includegraphics[width=0.75\textwidth]{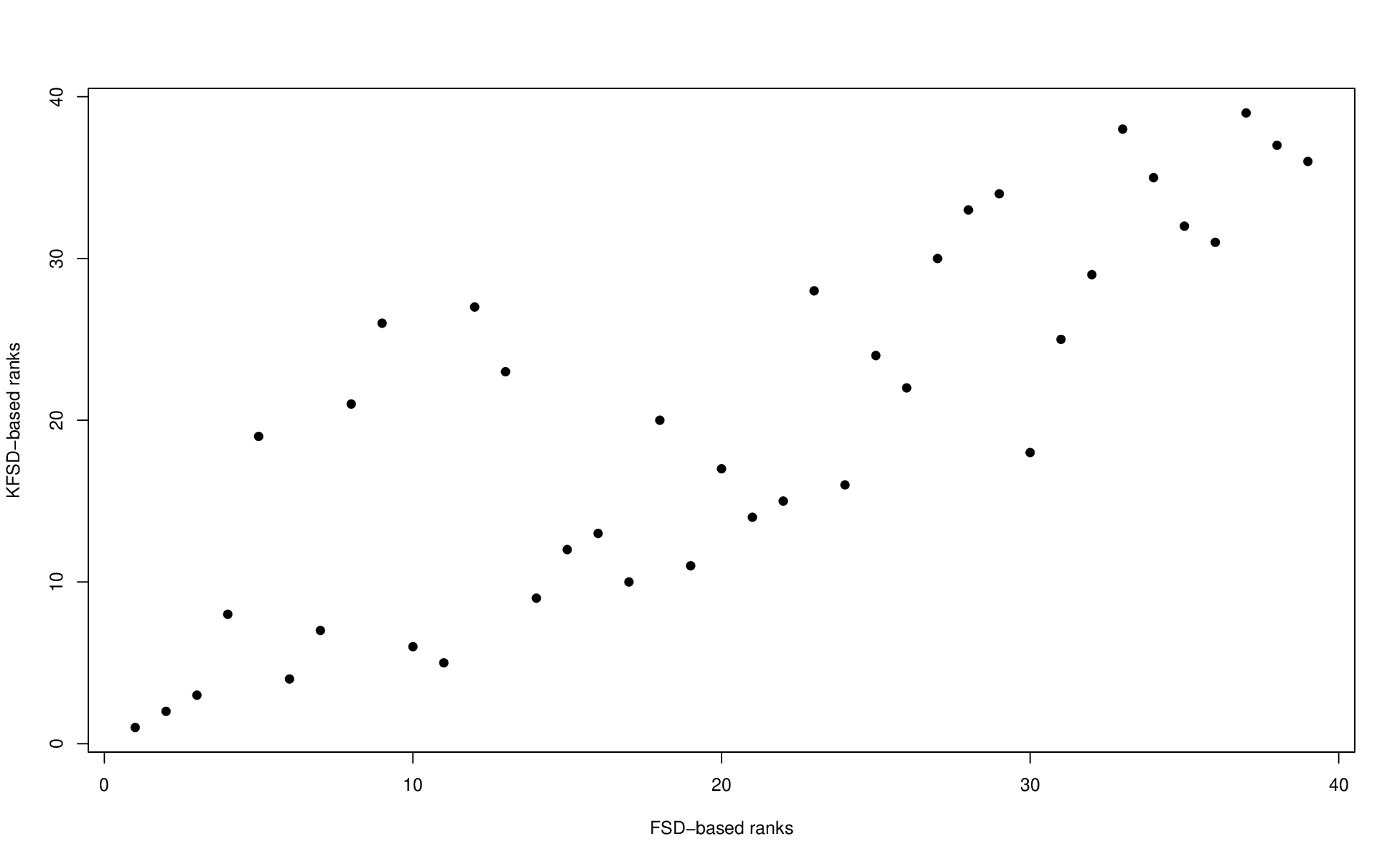}
\caption{Scatter plot of the FSD-based and KFSD-based ranks for the NO$_{x}$ data set.}
\label{fig:nox_RR}
\end{figure*}

Observing Figure \ref{fig:noxNW} we notice at least two features that can be described as complex and/or local: first, the data set contains NO$_{x}$ levels having a potential atypical behavior; second, the data set shows partial asymmetry, i.e., between roughly 10 and 24 hours there are many relatively low NO$_{x}$ levels and few relatively high NO$_{x}$ levels. Therefore, it seems interesting to compare the behavior of global and local functional depths using this sample affected by potential outliers and asymmetry.\\
\indent Figures \ref{fig:nox_Pairs} and \ref{fig:nox_RR} and Table \ref{tab:nox_Spe} mimic Figures \ref{fig:phonemes_Pairs} and \ref{fig:phonemes_RR} and Table \ref{tab:phonemes_Spe} for the NO$_{x}$ data set. When comparing all the depths between each other in Figure \ref{fig:nox_Pairs}, we see that the juxtaposition between global and local depths exists but it appears less strong than in the phonemes data set. 

However, analyzing Table \ref{tab:nox_Spe} we see that for each depth measure the highest Spearman's rank correlation coefficient is still observed with a depth measure of the same nature, and therefore global and local depths show different behaviors also when they are used to analyze a data set affected by the complex features identified in the NO$_{x}$ data set. 

\begin{table}[!htbp]
\caption{Spearman's rank correlation matrix of FSD, FMD, MBD, KFSD and HMD values for the NO$_{x}$ data set.}
\centering
\scalebox{1}{
\begin{tabular}{c|ccccc}
\hline
 & FSD & FMD & MBD & KFSD & HMD\\
\hline
FSD & 1.00 & 0.83 & 0.82 & 0.82 & 0.80\\
FMD & 0.83 & 1.00 & 0.97 & 0.75 & 0.73\\
MBD & 0.82 & 0.97 & 1.00 & 0.67 & 0.64\\
KFSD & 0.82 & 0.75 & 0.67 & 1.00 & 0.99\\
HMD & 0.80 & 0.73 & 0.64 & 0.99 & 1.00\\
\hline
\end{tabular}}
\label{tab:nox_Spe}
\end{table}

Focusing on FSD and KFSD, in Figure \ref{fig:nox_RR} we observe that these depths have a stronger relationship than in the phonemes data set. However, there are several observations for which the FSD-based ranks differ significantly from the KFSD-based ranks. For example, it is easily seen a group of five observations having FSD-based ranks roughly between 5 and 15 and KFSD-based ranks roughly between 20 and 30. In Section \ref{sec:rea} we give more details about the possible reasons why these observations may `benefit'' from the use of a local depth instead of a global one.

\section{Simulation study}
\label{sec:stu}

We have anticipated that global and local depths may behave differently when used to analyze real functional data sets, and in this section we present the results of a simulation study. The goal of the study is to establish when we should expect that a local functional depth may provide alternative data insights to the ones that would arise using a global functional depth.\\
\indent We are interested in models capable to replicate specific data features such as:

\begin{itemize}
\item absence of complex/local features;
\item presence of atypical observations;
\item asymmetry;
\item bimodality and presence of isolated observations.
\end{itemize}

\begin{figure*}[!htbp]
\centering
\includegraphics[width=0.75\textwidth]{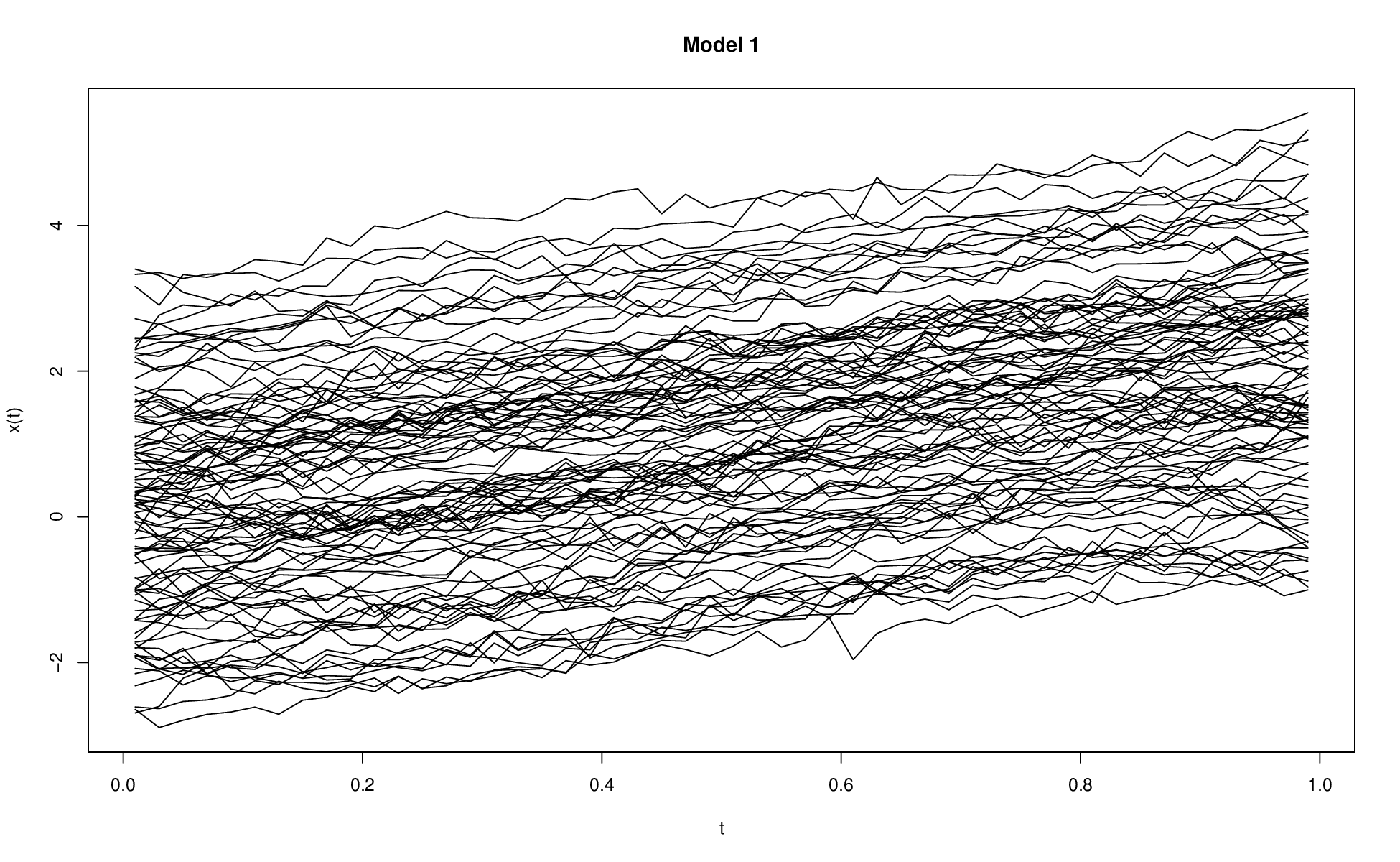}
\caption{Simulated data set from Model 1.}
\label{fig:model01}
\end{figure*}

To do this, we consider models based on truncated Karhunen-Lo\`eve expansions to which we add an error term. For example, the curves generating process defining the first model (Model 1) is given by

\begin{equation}
\label{eq:KL01}
x(t) = \mu(t) + \xi_{1}\phi_{1}(t) + \xi_{2}\phi_{2}(t) + \epsilon(t),
\end{equation}

\noindent where $t \in \left\{\frac{s-0.5}{50}, s = 1, \ldots, 50\right\}$, $\mu(t) = 2t$, $\xi_{1} \sim N(0, \lambda_{1})$ and $\lambda_{1} = 1.98$, $\xi_{2} \sim N(0, \lambda_{2})$ and $\lambda_{2} = 0.02$, $\phi_{1}(t) = 1$, $\phi_{2}(t) = \sqrt{7}\left(20t^3-30t^2+12t\right)$ and $\epsilon(t) \sim N(0, \sigma^2 = 0.01)$. Figure \ref{fig:model01} shows a simulated data set under Model 1 with sample size $n=100$. We use this sample size along the whole simulation study.

\begin{figure*}[!htbp]
\centering
\includegraphics[width=0.75\textwidth]{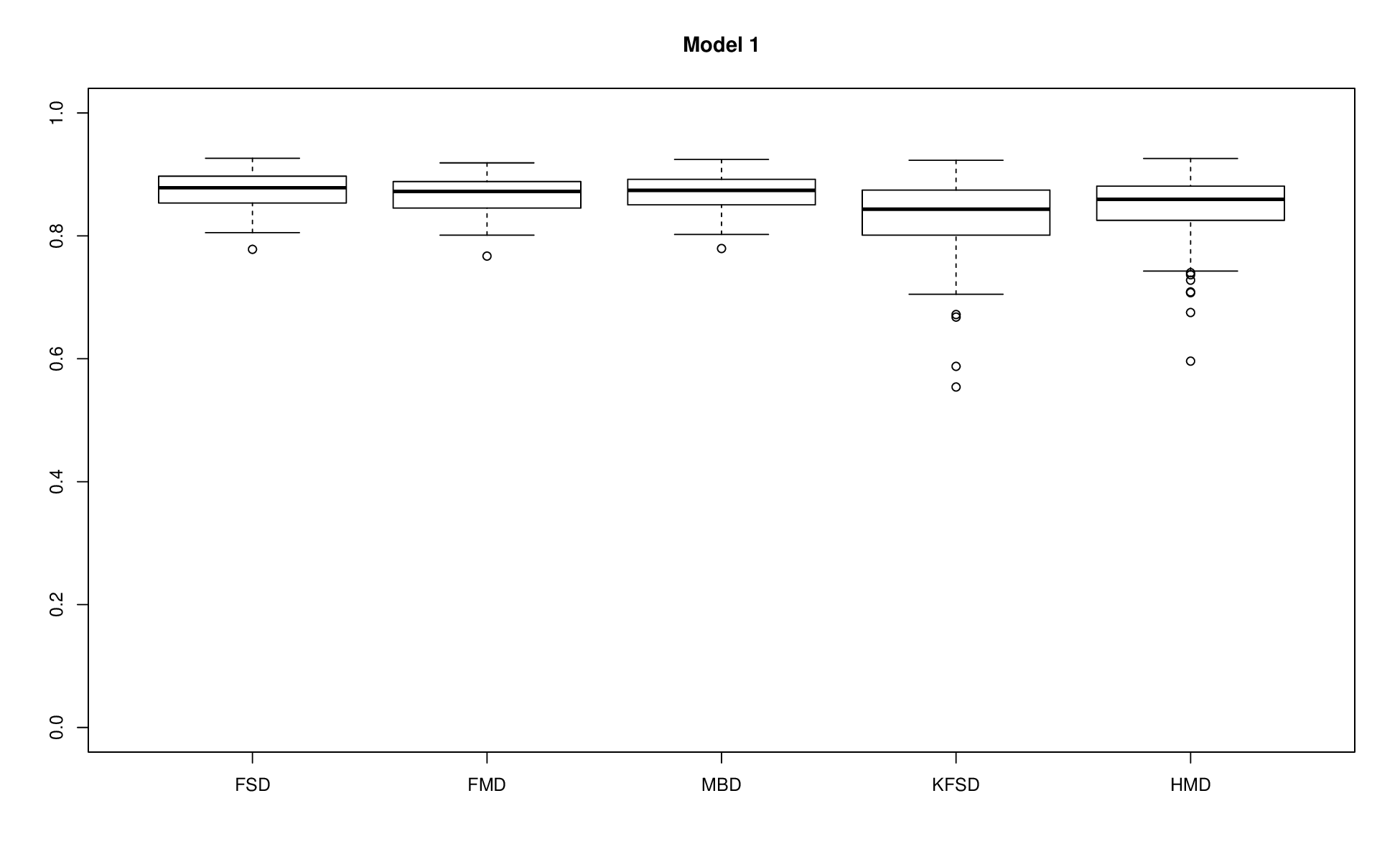}
\caption{Model 1: boxplots of the Spearman's rank correlation coefficients between the FSD, FMD, MBD, KFSD and HMD values and the $f(\xi_{1})$ values.}
\label{fig:model01Box}
\end{figure*}

Model 1 generates functional data that strongly depend on the realizations of $\xi_{1}$. Since $\xi_{1}$ is normal, Model 1 represents a scenario where complex data features are absent. Models 2, 3 and 4 will be defined modifying the distribution of $\xi_{1}$ and they will reproduce other data features of our interest. The design of our simulation study allows the attainment of two goals: first, the considered models will both replicate and isolate specific data features, and, second, the theoretical densities of the realizations of $\xi_{1}$, say $f(\xi_{1})$, will represent a natural benchmark to evaluate the performances of the functional depths under study.\\
\indent We generate $100$ samples from Model 1, and we evaluate the performance of a functional depth with each generated data set from Model 1 looking at the Spearman's rank correlation coefficient between depth and $f(\xi_{1})$ values. Figure \ref{fig:model01Box} shows the five boxplots obtained under Model 1, and they illustrate that in absence of complex features there are very mild differences in favor of global depths, which behave similarly among them. Local depths show similar but slightly more variable performances. 

\begin{figure*}[!htbp]
\centering
\includegraphics[width=0.75\textwidth]{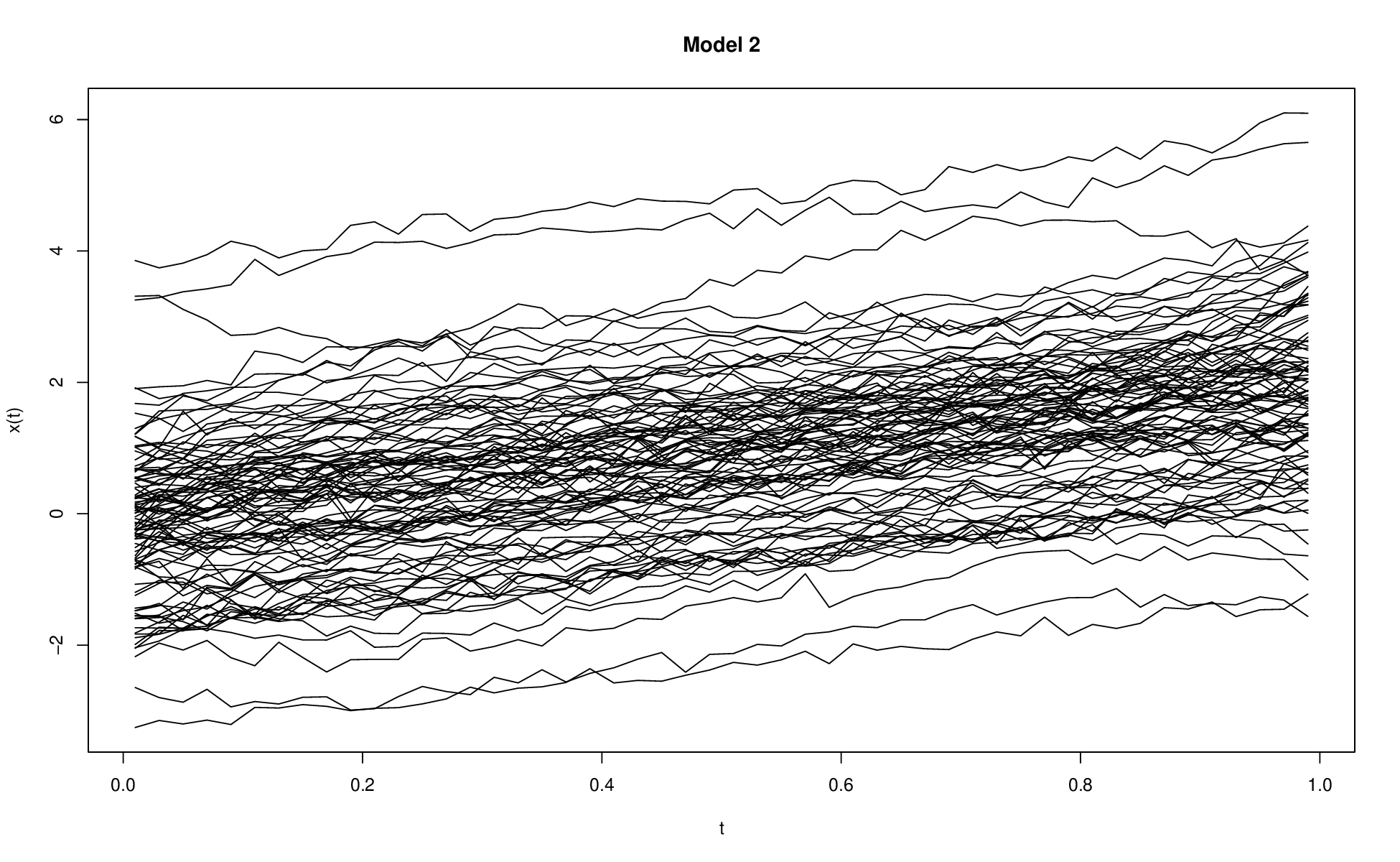}
\caption{Simulated data set from Model 2.}
\label{fig:model02}
\end{figure*}

Model 2 is obtained modifying the distribution of $\xi_{1}$:  under Model 2, $\xi_{1} \sim \sqrt{\lambda_{1}\frac{3}{5}}X$ and $X \sim t_{5}$. Note that this change allows us to obtain functional data sets potentially contaminated by atypical observations (see Figure \ref{fig:model02} for an example).

\begin{figure*}[!htbp]
\centering
\includegraphics[width=0.75\textwidth]{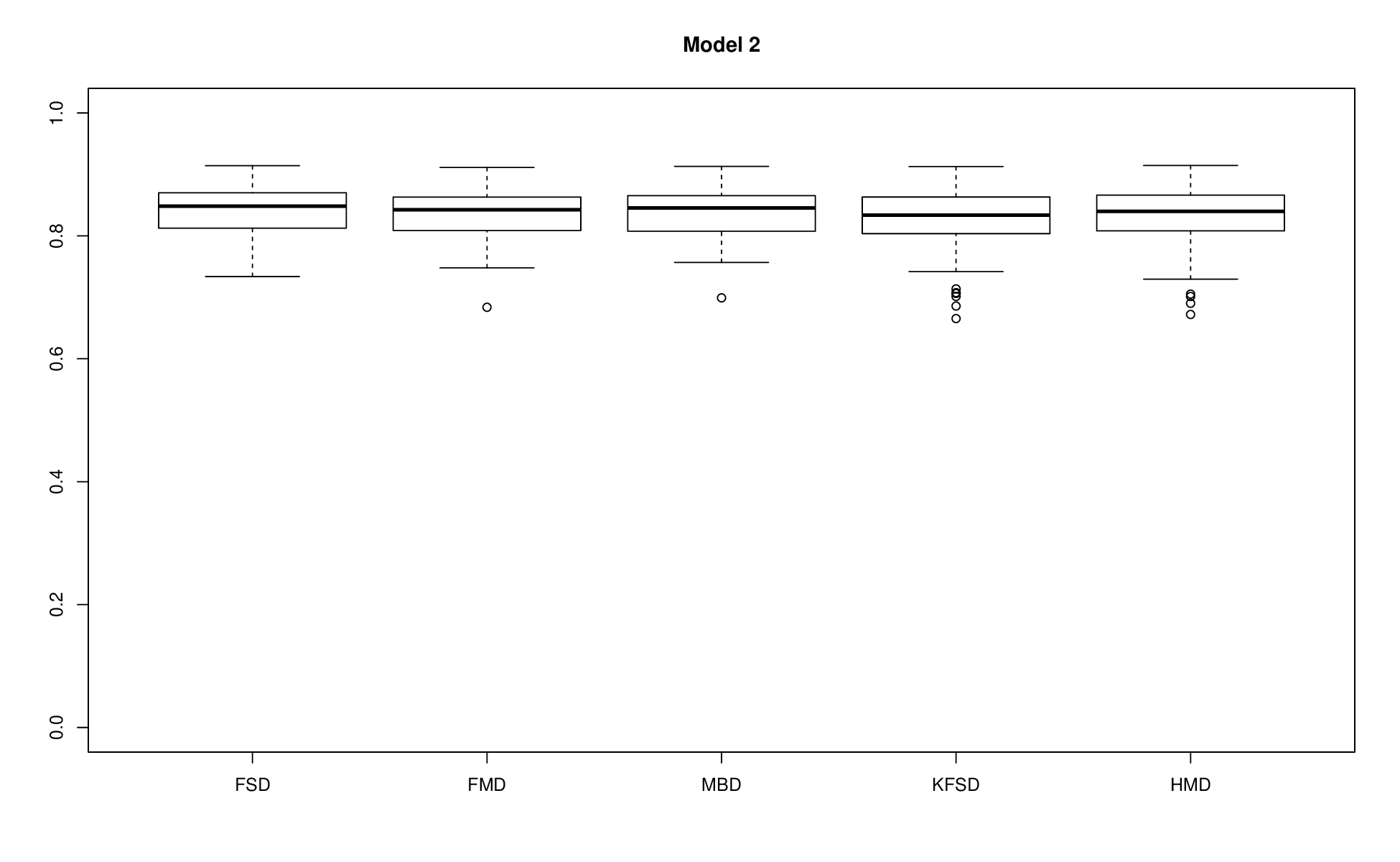}
\caption{Model 2: boxplots of the Spearman's rank correlation coefficients between the FSD, FMD, MBD, KFSD and HMD values and the $f(\xi_{1})$ values.}
\label{fig:model02Box}
\end{figure*}

Figure \ref{fig:model02Box} replicates Figure \ref{fig:model01Box} for Model 2, and, according to this new figure, in presence of a complex feature such as the existence of potential outliers both classes of depths behave very similarly. We claim that this result is due to the fact that both global and local depths analyze reasonably well those functional samples symmetrically contaminated by curves that are outlying because of their relative levels.

\begin{figure*}[!htbp]
\centering
\includegraphics[width=0.75\textwidth]{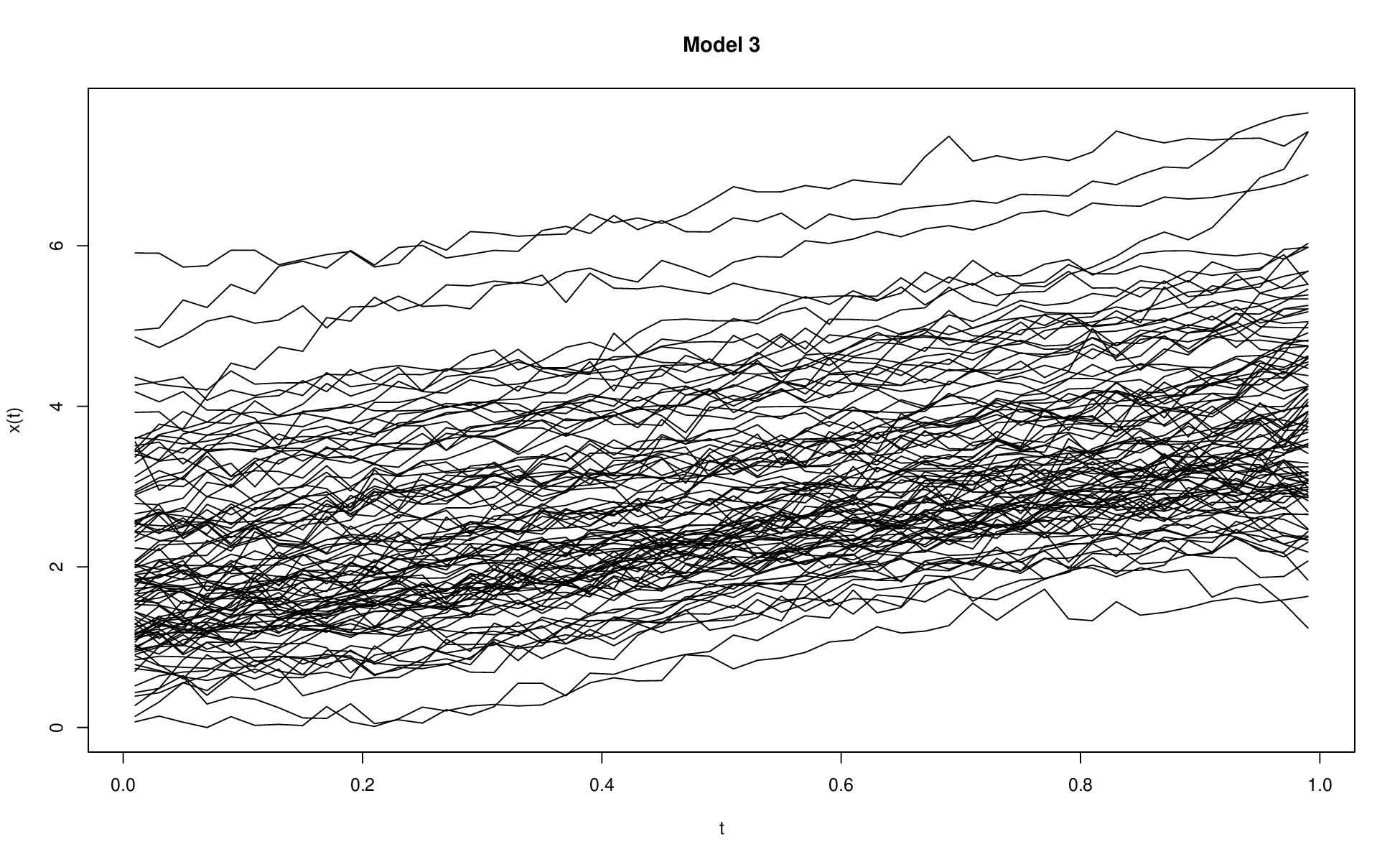}
\caption{Simulated data set from Model 3.}
\label{fig:model03}
\end{figure*}

To obtain Model 3 we consider an alternative modification of the distribution of $\xi_{1}$: under Model 3, $\xi_{1} \sim \sqrt{\lambda_{1}\frac{1}{10}}X$ and $X \sim \chi_{5}^{2}$. In this case the change allows to obtain asymmetric functional data sets, i.e., for all $t$, Model 3 generates many relatively low $x(t)$ and fewer relatively high $x(t)$ (see Figure \ref{fig:model03} for an example).

\begin{figure*}[!htbp]
\centering
\includegraphics[width=0.75\textwidth]{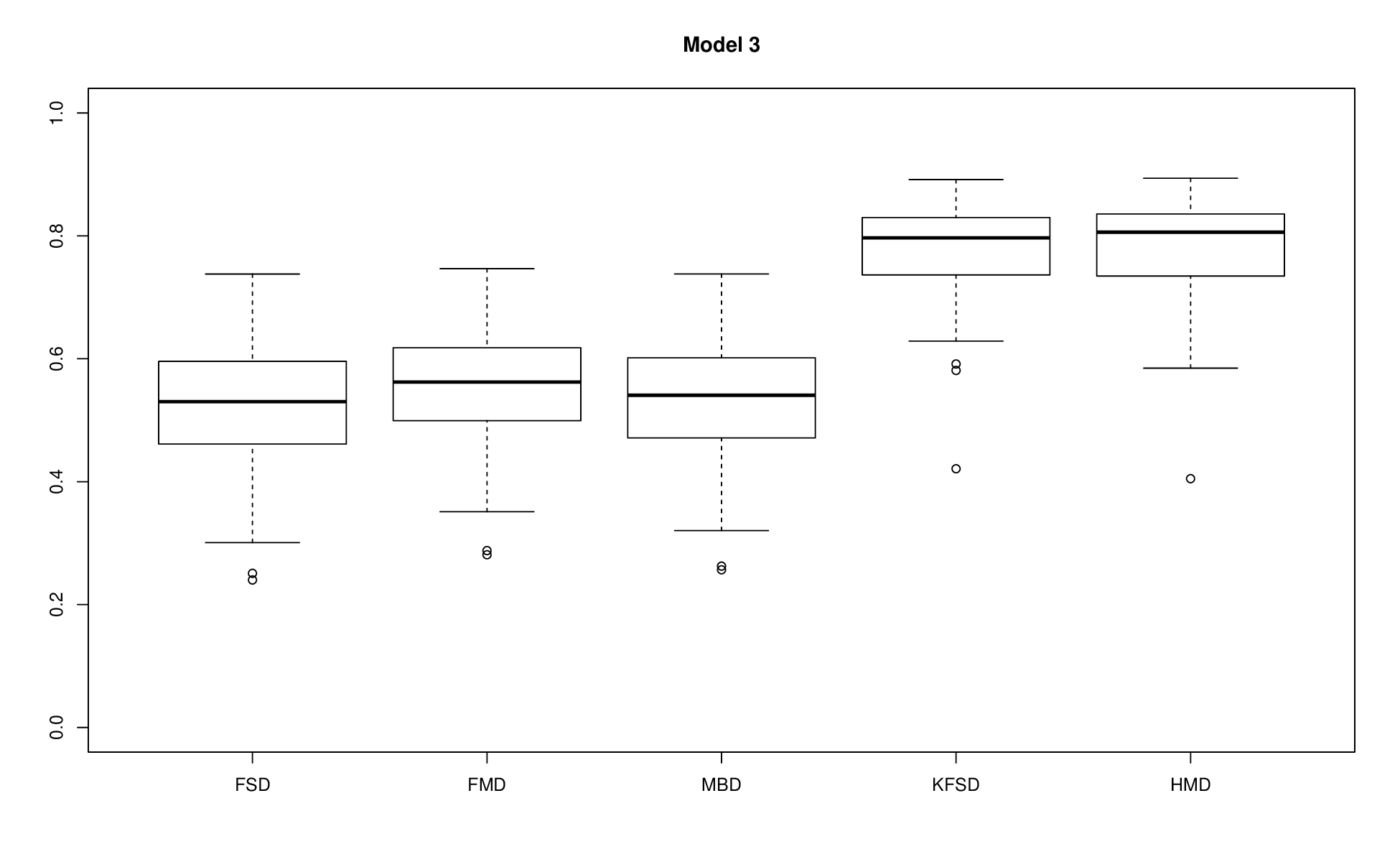}
\caption{Model 3: boxplots of the Spearman's rank correlation coefficients between the FSD, FMD, MBD, KFSD and HMD values and the $f(\xi_{1})$ values.}
\label{fig:model03Box}
\end{figure*}

Figure \ref{fig:model03Box} shows the boxplots obtained under Model 3, and it is clear that asymmetry represents a complex feature that affects the performances of all the depths under study, but it is handled much better by the local-oriented KFSD and HMD than by the global-oriented FSD, FMD and MBD.

\begin{figure*}[!htbp]
\centering
\includegraphics[width=0.75\textwidth]{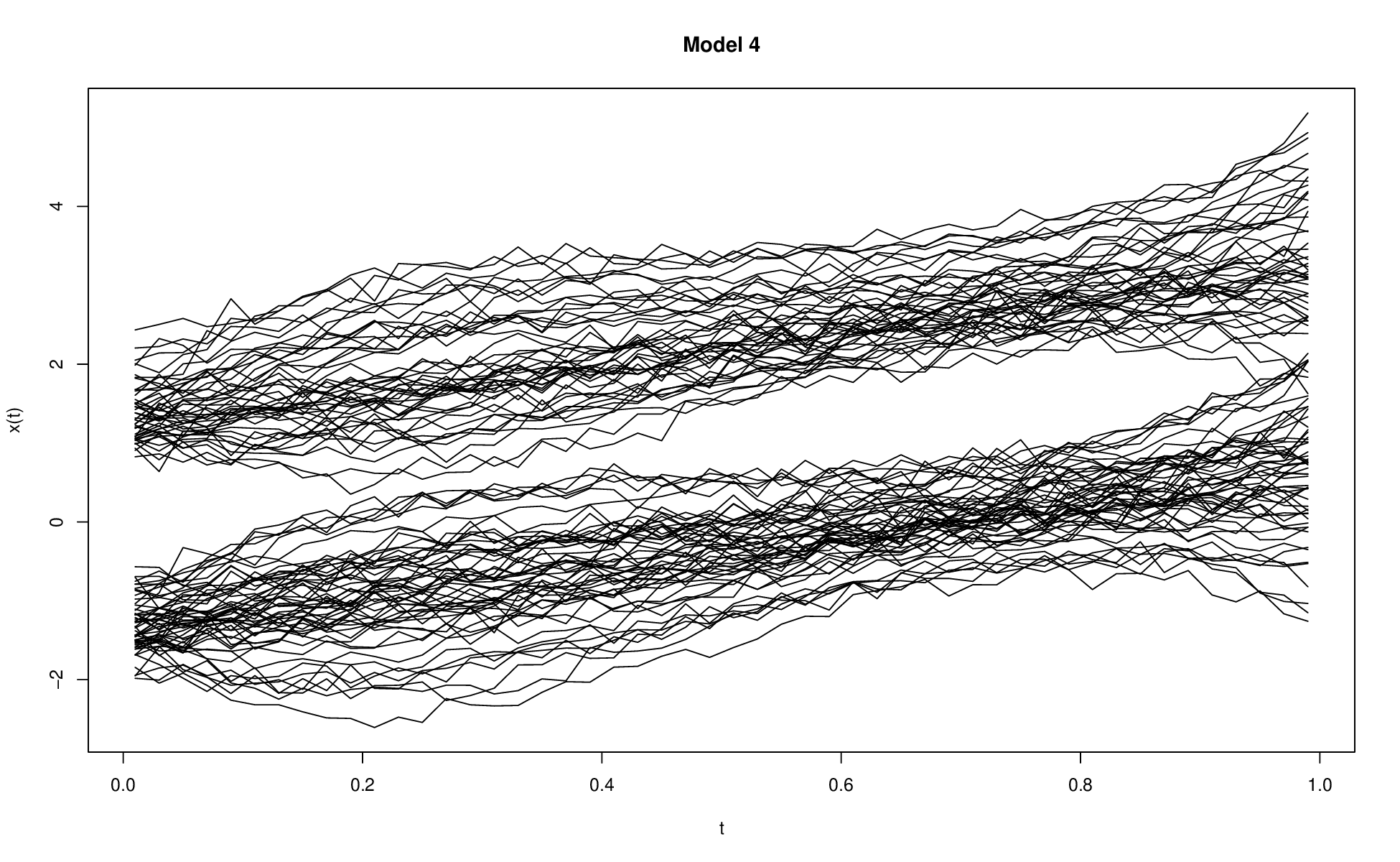}
\caption{Simulated data set from Model 4.}
\label{fig:model04}
\end{figure*}

Finally, we consider a mixture of two normals to obtain Model 4: with equal probability, $\xi_{1} \sim N\left(-\sqrt{\lambda_1-\frac{1}{10}}, \frac{1}{10}\right)$ or $\xi_{1} \sim N\left(\sqrt{\lambda_1-\frac{1}{10}}, \frac{1}{10}\right)$. We employ Model 4 to obtain data showing bimodality and potential presence of isolated observations lying between the two main groups of curves (see Figure \ref{fig:model04} for an example).
 
\begin{figure*}[!htbp]
\centering
\includegraphics[width=0.75\textwidth]{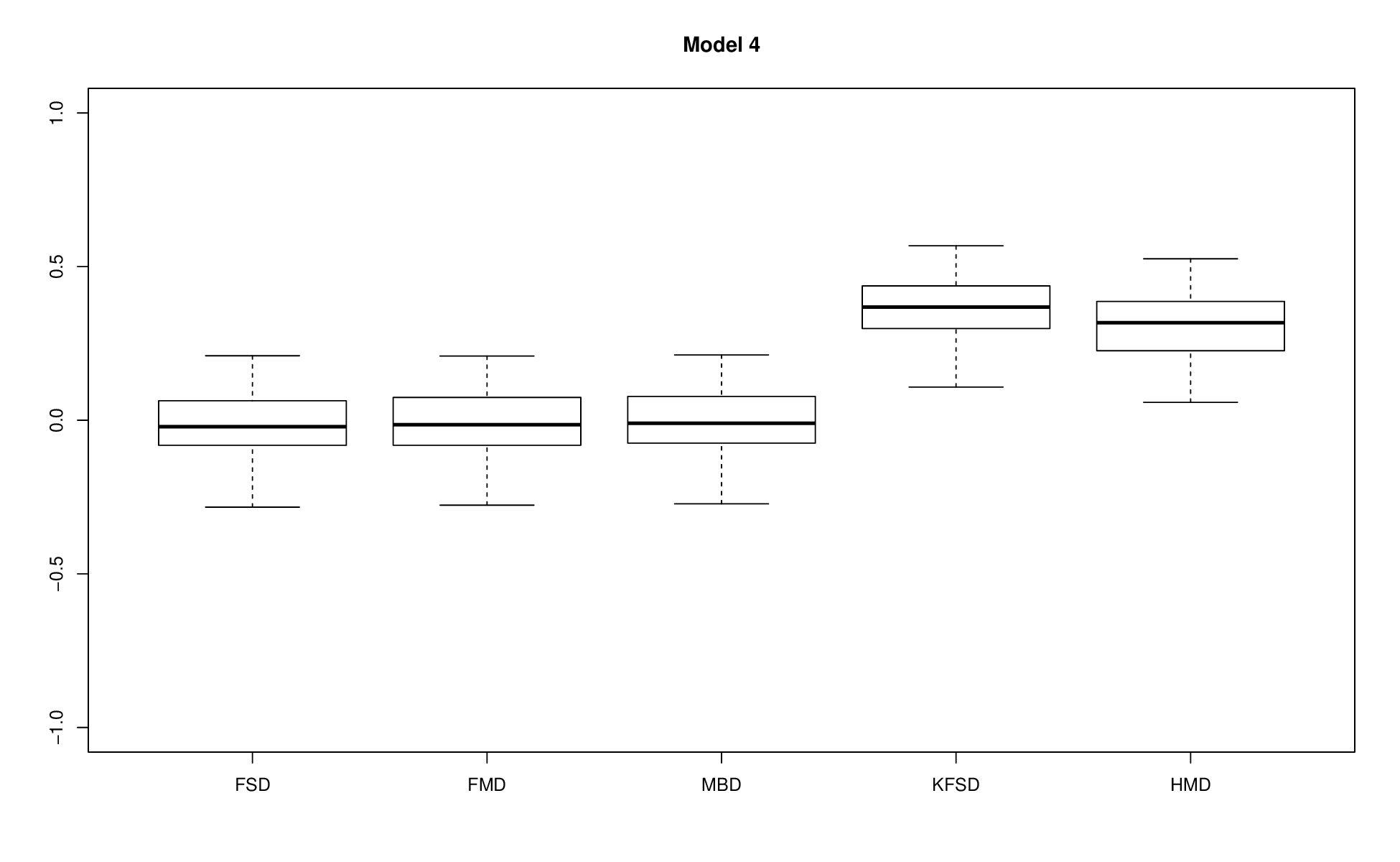}
\caption{Model 4: boxplots of the Spearman's rank correlation coefficients between the FSD, FMD, MBD, KFSD and HMD values and the $f(\xi_{1})$ values.}
\label{fig:model04Box}
\end{figure*}

Due to the behaviors of FSD, FMD and MBD under Model 4, when reporting the boxplots in Figure \ref{fig:model04Box}, we use $[-1, 1]$ as range for the vertical axis. The information provided by Figure \ref{fig:model04Box} suggests that the ranking of whole bimodal data sets represents a problem that is hard to be handled in an unsupervised way by the functional depths under study. However, the local-oriented KFSD and HMD show a generally positive association with the benchmark $f(\xi_{1})$, whereas for the global-oriented FSD, FMD and MBD we observe Spearman's rank correlation coefficients that vary symmetrically around 0.

The results of the simulation study presented in this section have shown that the behaviors of global and local functional depths can be fairly similar under some circumstances (e.g., absence of complex data features and presence of a particular type of outliers), but quite different under others (e.g., asymmetry and bimodality). Before drawing our conclusions, in the next section we link the results of this section with the functional depth analyses of the phonemes and NO$_{x}$ data sets introduced in Section \ref{sec:motExa}.

\section{Final remarks on the motivating examples}
\label{sec:rea}

\begin{figure*}[!htbp]
\centering
\includegraphics[width=0.75\textwidth]{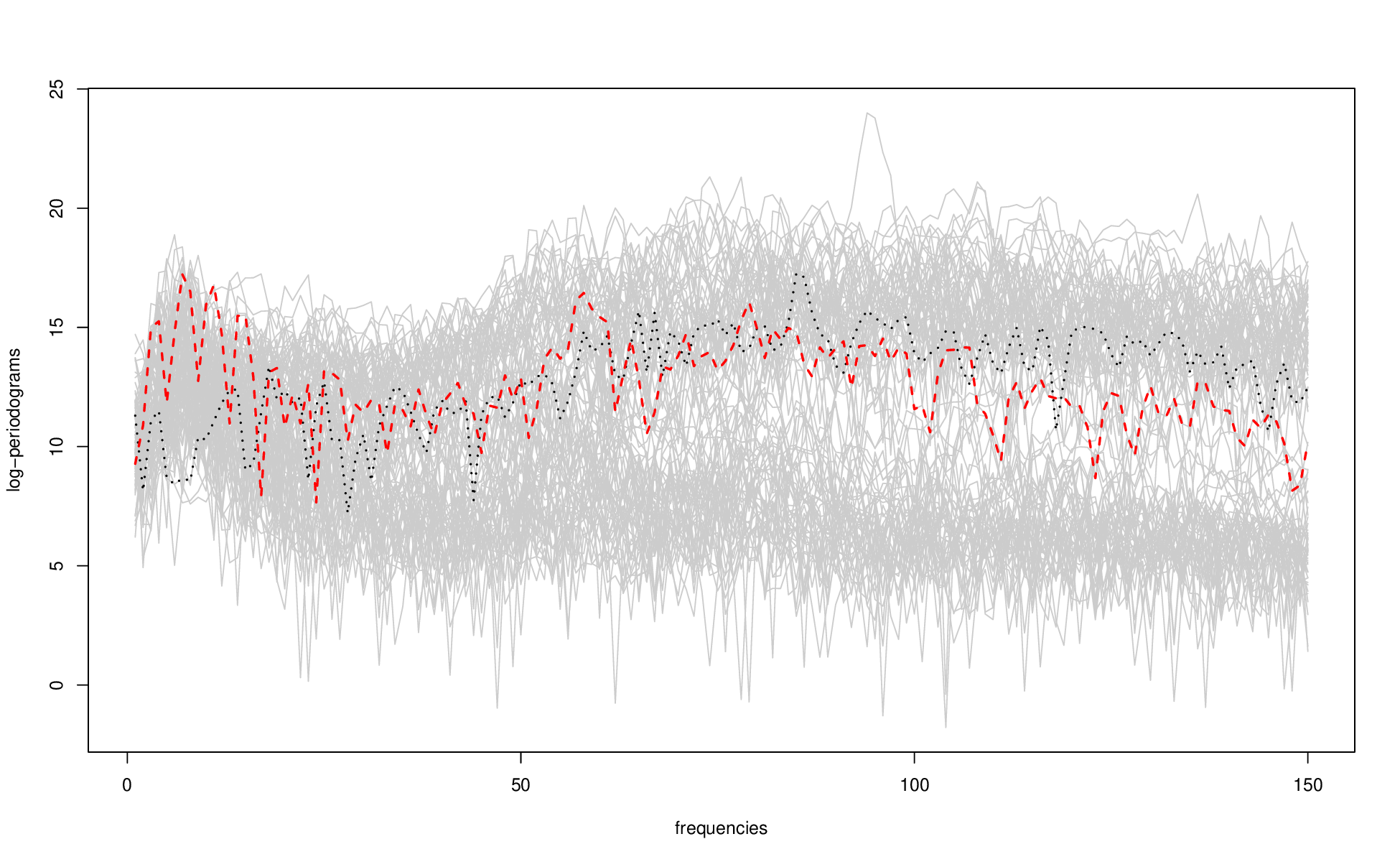}
\caption{Phonemes data set. The dotted curve has rank equal to 99 and 94 according to FSD and KFSD, respectively. The dashed curve has rank equal to 100 and 26 according to FSD and KFSD, respectively.}
\label{fig:phonemesHL}
\end{figure*}

In Section \ref{ssec:motExa02} we have shown that global and local depths produce different depth analyses for the phonemes data set ($Y_{ph}$). In that section and in Section \ref{sec:stu} we have explained that this is due to the fact that global and local depths handle differently bimodal data sets. Using FSD as global depth and KFSD as local depth, we highlight two examples of two different types of phoneme curves: an observation whose depth analysis in practice is not affected by the choice of the depth measure, and another observation for which the opposite occurs. The first curve is ranked as the 99th and 94th curve by FSD and KFSD, respectively. The second curve is ranked as the 100th and 26th curve by FSD and KFSD, respectively. Both observations are easily identifiable in Figure \ref{fig:phonemes_RR}, while in Figure \ref{fig:phonemesHL} they are presented in their functional form, together with the whole data set.

\begin{figure*}[!htbp]
\centering
\includegraphics[width=0.75\textwidth]{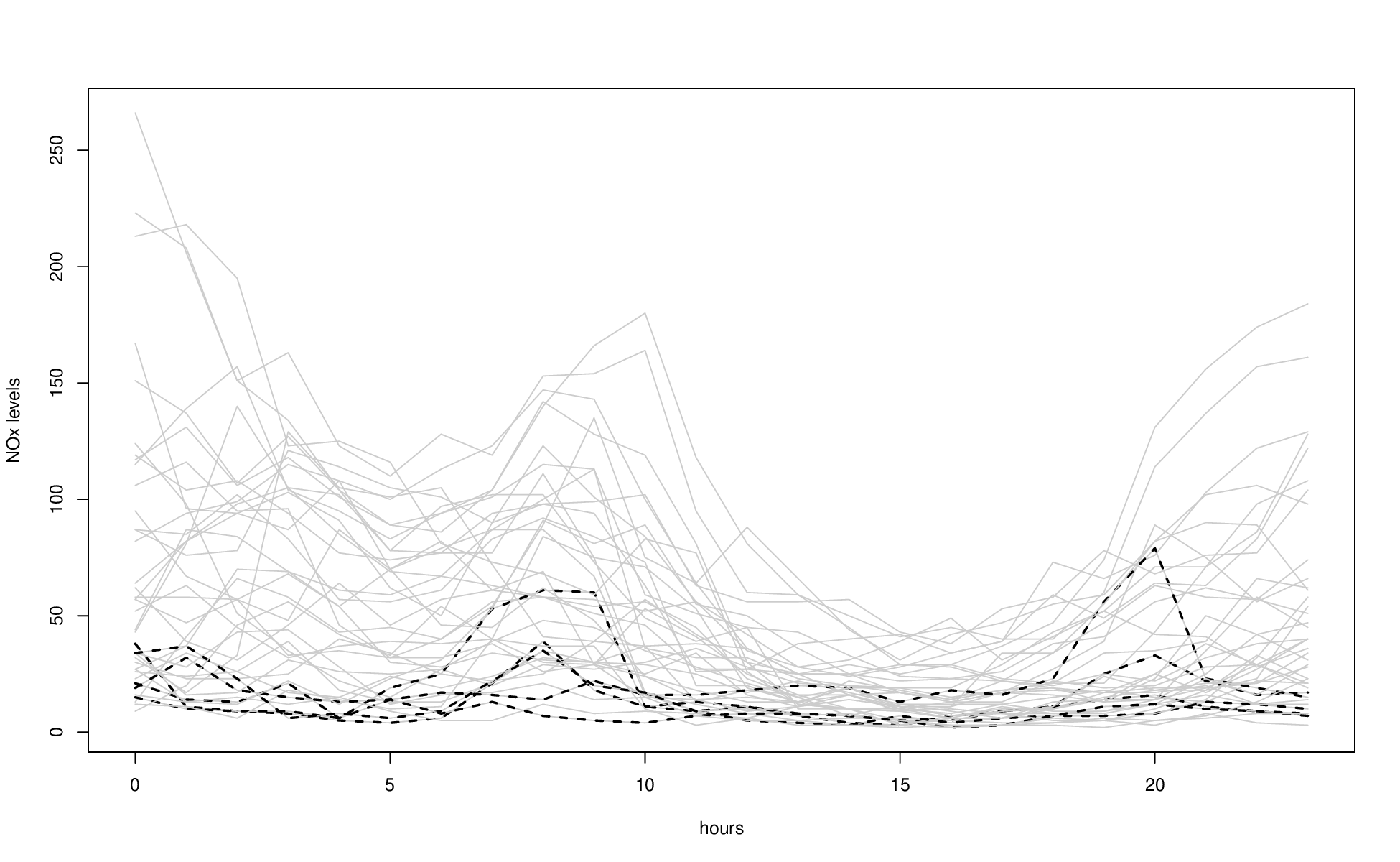}
\caption{NO$_{x}$ data set. The dashed curves have FSD-based ranks roughly between 5 and 15 and KFSD-based ranks roughly between 20 and 30.}
\label{fig:noxHL}
\end{figure*}

The dotted curve in Figure \ref{fig:phonemesHL} is a \textit{sh} phoneme that seems fairly representative of its class. This curve is among the deepest curves of the whole phoneme data set  according to FSD and KFSD, and it represents an example of coincidence between a global and a local depth analysis. An example of discordance is given by the dashed curve in Figure \ref{fig:phonemesHL}, which is a \textit{dcl} phoneme that seems potentially atypical and relatively isolated with respect to the whole phoneme data set. This second curve is the deepest curve according to FSD, but it is only the 74th deepest curve according to KFSD. This result illustrates that a local depth tends to penalize isolated observations, and that therefore it should handle better the presence of bimodality in the data.\\
\indent The FSD and KFSD analyses of the NO$_{x}$ data set presented in Section \ref{ssec:motExa01} show less differences than what we have observed in the phonemes data set. However, we have already reported the presence of a group of five observations with KFSD-based ranks higher than their corresponding FSD-based ranks (see Figure \ref{fig:nox_RR} and our remarks on it in Section \ref{ssec:motExa01}). In Figure \ref{fig:noxHL} we present this group of curves in their functional form, together with the rest of the NO$_{x}$ observations.

Recall that the NO$_{x}$ data set under analysis refers to nonworking days, and that for this reason we observe relatively many curves with low NO$_{x}$ levels and fewer with high NO$_{x}$ levels. In other words, the data set shows asymmetry (see Figure \ref{fig:noxNW}). All the dashed curves in Figure \ref{fig:noxHL} have low NO$_{x}$ levels, at least during a period of the day, and therefore they have a behavior that many other curves show. However, the global approach behind FSD penalizes perhaps excessively these observations for having low NO$_{x}$ levels, while the local approach behind KFSD is able to recognize that such low values are a common feature in the functional sample.

\section{Conclusions}
\label{sec:con}
The differences between a global and a local approach to the depth problem have been investigated in the multivariate framework. With the aim of extending this knowledge to the functional framework, in this paper we have studied and compared the behavior of three global and two local functional depths. We have illustrated that local functional depths may behave differently with respect to their global alternatives. Indeed, using real and simulated data sets, we have observed that analyses based on local depths may be more appropriate under specific scenarios. In this work we have identified at least two: first, in presence of asymmetry (see Model 3 and NO$_{x}$ analyses); second, in presence of bimodality and isolated observations (see Model 4 and phonemes analyses).

\bibliographystyle{spbasic}      
\bibliography{refsPostThesis} 

\end{document}